\documentclass[authoryear,final,5p,times,twocolumn]{elsarticle}

\usepackage{graphics}

\usepackage{amssymb}
\usepackage{hyperref}

\journal{Advances in Space Research}

\begin{document}

\begin{frontmatter}



\title{Local Luminous Infrared Galaxies: Spatially resolved 
mid-infrared observations with {\it Spitzer}/IRS 
\tnoteref{label1}}
\tnotetext[label1]{Based on observations obtained with the Spitzer Space Telescope, which is operated by the Jet Propulsion Laboratory, California Institute of Technology, under NASA contract 1407}


\author{Almudena Alonso-Herrero\fnref{label2,label3}}
\author{Miguel Pereira-Santaella\fnref{label2}}
\author{George H. Rieke\fnref{label3}} 
\author{Luis Colina\fnref{label2}}
\author{Charles W. Engelbracht\fnref{label3}}
\author{Pablo G. P\'erez-Gonz\'alez\fnref{label4,label3}}
\author{Tanio D\'{\i}az-Santos\fnref{label2,label6}}
\author{J.-D. T. Smith\fnref{label5}}

\fntext[label2]{Instituto de Estructura de la Materia, CSIC, Serrano 121, 
E-28006, Madrid, Spain}

\fntext[label3]{Steward Observatory, University of Arizona, 933
  N. Cherry Avenue,  Tucson, AZ 85721, USA}

\fntext[label4]{Departamento de Astrof\'{\i}sica, Facultad de
  CC. F\'{\i}sicas, Universidad
  Complutense de Madrid, Avenida Complutense S/N, E-28040, Madrid, Spain}
\fntext[label5]{Ritter Astrophysical Research Center, University of
  Toledo, 2801 West Bancroft Street, Toledo, OH 43603, USA}

\fntext[label6]{Current address: University of Crete, Department of Physics,
P.O. Box 2208, GR-71003, Heraklion, Greece}

\begin{abstract}
Luminous Infrared (IR) Galaxies (LIRGs, $L_{\rm IR}=10^{11}-10^{12}$L$_\odot$)
  are an important
cosmological class of galaxies as they are the main contributors to the
co-moving star formation rate density of the universe at $z=1$. 
In this paper we present a guaranteed time observation (GTO) {\it Spitzer} 
InfraRed Spectrograph (IRS) program aimed to obtain 
spectral mapping of  a sample of 14 local ($d<76\,$Mpc) LIRGs. The 
data cubes map, {\rm at least}, 
the central $20\,{\rm arcsec} \times 20\,{\rm arcsec}$ 
to $30\,{\rm arcsec} \times 30\,{\rm arcsec}$  regions of the galaxies,
and use all four IRS 
modules covering the full $5-38\,\mu$m spectral range. 
The final goal of this
project is to characterize fully the mid-IR properties of local LIRGs as a first
step to understanding their more distant counterparts.  
In this paper we present the first results of this GTO program. 
The IRS spectral mapping data allow us to build spectral maps of the
bright mid-IR emission 
lines (e.g., [Ne\,{\sc ii}]$12.81\,\mu$m, [Ne\,{\sc iii}]$15.56\,\mu$m, 
[S\,{\sc iii}]$18.71\,\mu$m, H$_2$ at $17\,\mu$m),  
continuum, the 6.2 and $11.3\,\mu$m 
polycyclic aromatic hydrocarbon (PAH) features, and the $9.7\,\mu$m
silicate feature,  
as well as to extract 1D spectra for regions of
interest in each galaxy. The IRS data are used to obtain spatially
resolved measurements of the extinction using the $9.7\,\mu$m silicate
feature, and to trace star forming regions using the
neon lines and the PAH features. 
We also investigate a number of
active galactic nuclei (AGN) indicators, 
including the presence of  
high excitation emission lines and 
a strong dust continuum emission at around $6\,\mu$m. 
We finally use the integrated {\it Spitzer}/IRS spectra as 
templates of local LIRGs. We discuss several possible uses for
these templates, including the calibration of the star formation rate of 
IR-bright galaxies at high redshift. We also  predict the 
intensities of the brightest
mid-IR emission lines for LIRGs as a function of redshift, 
and compare them with the expected sensitivities
of future space IR missions.

\end{abstract}


\end{frontmatter}


\section{Introduction}

The importance of infrared (IR) bright galaxies 
has been increasingly appreciated since their discovery more than 30 years ago
(Rieke  \& Low 1972), and the detection of large numbers  by  {\it IRAS} (Soifer et
al. 1987). Although the Ultraluminous Infrared Galaxies (ULIRGs, 
with IR luminosities $8-1000\,\mu$m, 
$L_{\rm IR} = 10^{12}$ to $10^{13}\,{\rm  L}_\odot$)
get much of the attention because they are so dramatic,
Luminous Infrared Galaxies (LIRGs, $L_{\rm IR} = 10^{11}$ to $10^{12}\,{\rm
  L}_\odot$) are much more common, accounting for $\sim$ 5\% of the local
IR background compared with $<$ 1\% for the ULIRGs (Lagache et al. 2005). 
They are of intrinsic interest because they provide
insight to star formation and nuclear activity under extreme conditions 
and should also include former ULIRGs where the star formation is
dying off (see Murphy et al. 2001). 
They take part in the controversy over
the formation of AGN and its relation to star formation and high
levels of IR emission (see review by Sanders \& Mirabel 1996).

Local LIRGs may also be prototypes for forming galaxies 
at high redshift (Lagache et al. 2005). 
Deep {\it Spitzer} detections at $24\,\mu$m are dominated by 
LIRGs and ULIRGs. LIRGs are the main contributors to the co-moving star
formation rate (SFR) density 
in the $z\sim 1-2$ range (Elbaz et al. 2002; Le Floc'h et al. 
2005; P\'erez-Gonz\'alez et al. 2005; Caputi et al. 2007), 
and contribute nearly 50\% of the cosmic
IR background at  $ z \sim 1$  (Lagache et al. 2005). Moreover, the
mid-IR spectra of high redshift  ($z\sim 2)$ very luminous IR galaxies
($L_{\rm IR}>10^{12}$L$_\odot$) are more similar to those of local
starbursts and LIRGs (see e.g., Farrah et al. 2008; Rigby et
al. 2008; Alonso-Herrero et al. 2009a) than those of local ULIRGs. 
This may just reflect the fact that at high-$z$ 
star-formation was taking place over a few kiloparsec scales rather than 
in very compact ($<$1kpc) regions as  is the case for 
local ULIRGs (e.g., Soifer et al. 2000).

\begin{table*}
\begin{center}
\caption{The sample of local LIRGs mapped with the IRS.}

\begin{tabular}{llcclc}

\hline
Galaxy Name  &
IRAS Name &
$v_{\rm hel}$ &  
Dist & $\log L_{\rm IR}$ & 
Spect. \\
 & & (km s$^{-1}$) & 
(Mpc) & (L$_\odot$) & class\\
(1) & (2) & (3) &(4) & (5) & (6) \\
\hline

NGC~2369           & IRASF~07160$-$6215 & 3237  & 44.0 & 11.10  &--    \\
NGC~3110           & IRASF~10015$-$0614 & 5034  & 73.5 & 11.31: &H\,{\sc ii}\\
NGC~3256$^{*}$           & IRASF~10257$-$4339 & 2814  & 35.4 & 11.56  &H\,{\sc ii}\\
Arp~299$^{**}$       & IRASF~11257+5850 & 3121  & 47.7 & 11.88
&H\,{\sc ii}, Sy2\\
ESO~320-G030       & IRASF~11506$-$3851 & 3232  & 37.7 & 11.10  &H\,{\sc ii}   \\
NGC~5135           & IRASF~13229$-$2934 & 4112  & 52.2 & 11.17  &Sy2   \\
Zw~049.057         & IRASF~15107+0724 & 3897  & 59.1 & 11.27: &H\,{\sc
  ii} \\
---                & IRASF~17138$-$1017 & 5197  & 75.8 & 11.42  &H\,{\sc ii}   \\
IC~4687/IC~4686$^{***}$ & IRASF~18093$-$5744 & 5200/4948  & 74.1 & 11.55:  &H\,{\sc ii} (both)\\ 
NGC~6701           & IRASF~18425+6036 & 3965  & 56.6 & 11.05  & Composite    \\
NGC~7130           & IRASF~21453$-$3511 & 4842  & 66.0 & 11.35  &Sy/L  \\
IC~5179            & IRASF~22132$-$3705 & 3422  & 46.7 & 11.16  &H\,{\sc ii}\\
NGC~7591           & IRASF~23157+0618 & 4956  & 65.5 & 11.05  & Composite    \\
NGC~7771           & IRASF~23488+1949 & 4277  & 57.1 & 11.34   &H\,{\sc ii}\\
\hline
\end{tabular}

Notes. --- Column~(1): Galaxy name. $^{*}$NGC~3256 is a merger system
with two nuclei, refered to as north and south. $^{**}$Arp~299 is composed of 
IC~694 (eastern component) and  NGC~3690 (western component).
 $^{***}$Only IC~4687 was observed in spectral mapping mode. 
Column~(2): {\it IRAS} denomination
  from Sanders et al. (2003). Column~(3): 
Heliocentric velocity from NED. Column~(4): Distance taken from
Sanders et al. (2003) assuming $H_0=75\,{\rm km
  \,s}^{-1}\,{\rm Mpc}^{-1}$. Column~(5): $8-1000\,\mu$m IR
luminosity taken from Sanders et al. (2003), where the suffix ``:'' means
large uncertainty (see Sanders et al. 2003 for details). 
$^{*}$The IR luminosity is for the system Arp~299=IC~694+NGC~3690. 
Column~(6): Nuclear activity class 
from optical spectroscopy. ``Sy''=Seyfert, 
``Composite''=intermediate between LINER and H\,{\sc ii} (see
  Alonso-Herrero et al. 2009b), ``H\,{\sc ii}''=H\,{\sc ii}
region-like, ``--''=no classification
available. $^{*}$The classifications for Arp~299 are: 
nuclear region of IC~694, or source A, is H\,{\sc 
  ii}-like, and the nuclear region of NGC~3690, or source B1, is a 
Sy2. The references for the nuclear class are given by
Alonso-Herrero et al. (2006a), except for the nuclear region of
NGC~3690 which is from Garc\'{\i}a-Mar\'{\i}n et al. (2006), and
NGC~6701 and NGC~7591, which are 
reported by  
Alonso-Herrero et al. (2009b). 
\end{center}
\end{table*}

Much of our knowledge of the mid-IR spectroscopic properties of 
local IR-bright
galaxies comes from {\it ISO} (e.g., Genzel et al. 1998; Rigopoulou et
al. 1999;  
Tran et al. 2001) and early results (Armus et al. 2004, 2007) 
with the InfraRed Spectrograph (IRS, Houck et
al. 2004) on {\it Spitzer}.  However, the majority of the {\it ISO}
works focused on  samples of local IR-bright galaxies or starburst
galaxies,  
and only include a few luminous LIRGs ($\log L_{\rm IR} \simeq 11.80\,{\rm
  L}_\odot$, see e.g., Genzel et al. 1998; Verma et al. 2003).  The
majority of the recent {\it Spitzer} results are concentrating on 
ULIRGs (e.g., Armus et al. 2007; Farrah et al. 2007), starburst
galaxies  (Brandl et al. 2006), and nearby star-forming galaxies (Dale
et al. 2006; Smith et al. 2007a). Recently, Armus et al. (2009)
  have presented the Great Observatories All-Sky LIRG Survey (GOALS) 
project which combines multiwavelength data, including {\it Spitzer}, of over
200 low-redshift ($z < 0.088$) LIRGs.

In this paper we summarize the goals and present the  
first results of a {\it Spitzer}/IRS 
program intended to obtain mid-IR ($5-38\,\mu$m) 
spectral mapping 
of a representative sample of fourteen local LIRGs (Table~1) selected
from the volume-limited sample 
of Alonso-Herrero et al. (2006a). It is only by quantifying the
integrated spectroscopic properties of a representative sample of
local LIRGs that we can interpret measurements at  $z \sim 1$, which
apply to the
integrated galaxy. The final goal of this
project is to characterize fully the mid-IR properties of local LIRGs as a first
step to understanding their more distant counterparts.  
The complete results from this LIRG program will be presented in a number of
forthcoming papers (Pereira-Santaella et al. 2009a, and 2009b in
preparation). 
The paper is organized as follows. Section~2 presents the sample,
observations and data reduction. Section~3 characterizes the
obscuration of LIRGs and Section~4 discusses the AGN and star
formation properties of the sample. Section~5 presents the integrated
mid-IR spectra and their potential use for studies of IR-bright
galaxies at high-$z$. Finally Section~6 gives a summary of this work.

\section{Sample, Spitzer/IRS Spectral Mapping Observations, and Data Reduction}

The galaxies chosen for this study are taken from a representative 
and volume-limited sample of LIRGs  originally drawn from 
the {\it IRAS} Revised Bright Galaxy Sample (Sanders et
al. 2003) for Pa$\alpha$ imaging (using a narrow-band filter) 
with NICMOS on the {\it HST} (Alonso-Herrero et al. 2006a). 
Hence  both the NICMOS sample and the sub-sample selected for IRS  
observations are restricted in redshift ($d<76\,$Mpc). 
For this IRS program we chose  LIRGs in
this sample with extended Pa$\alpha$ and mid-IR emission 
(Alonso-Herrero et al. 2006a and D\'{\i}az-Santos et al. 2008) 
so we could map out most of the IR emission of these galaxies. This is 
important for comparison with IR-selected high-z galaxies where the
entire galaxy 
is encompassed in the IRS slit.

As can be seen from Table~1, the IRS sample covers well the
range of IR luminosity of LIRGs (see Sanders et al. 2003), and the
nuclear activity classes of LIRGs. For this sample, $\sim 30$\%
  are classified as AGN, if we include ¨composite¨ objects, that is,
  nuclei classified as 
intermediate between H\,{\sc ii} and LINER using optical
  line  ratios (see Alonso-Herrero et
  al. 2009b). This fraction is similar to that found
  for LIRGs by 
Veilleux et al. (1995). The IR luminosities of this sample imply
SFRs of between 19 and $130\,{\rm M}_\odot\,{\rm yr}^{-1}$, using the
Kennicutt (1998) prescription. In Alonso-Herrero et al. (2009a) we recently
presented a detailed study of Arp~299. 

We obtained IRS mapping observations of the LIRG sample using 
 all four modules: Short-High (SH; $9.9-19.6\,\mu$m), 
Long-High (LH; $18.7-37.2\,\mu$m), Short-Low
(SL1; $7.4-14.5\,\mu$m  and SL2; $5.2-7.7\,\mu$m) 
and Long-Low (LL1; $19.5-38\,\mu$m and LL2; $14.0-21.3\,\mu$m). 
The plate scales are 1.85, 2.26, 4.46, and 5.08 arcsecond per pixel
for the SL, SH, LH,
and LL modules, respectively.
The IRS high spectral resolution data (SH and LH 
 modules) have $R\sim 600$, whereas the low resolution 
spectra (SL and LL) have $R\sim 60-126$. 

\begin{figure}
\includegraphics[width=8.5cm,angle=-90]{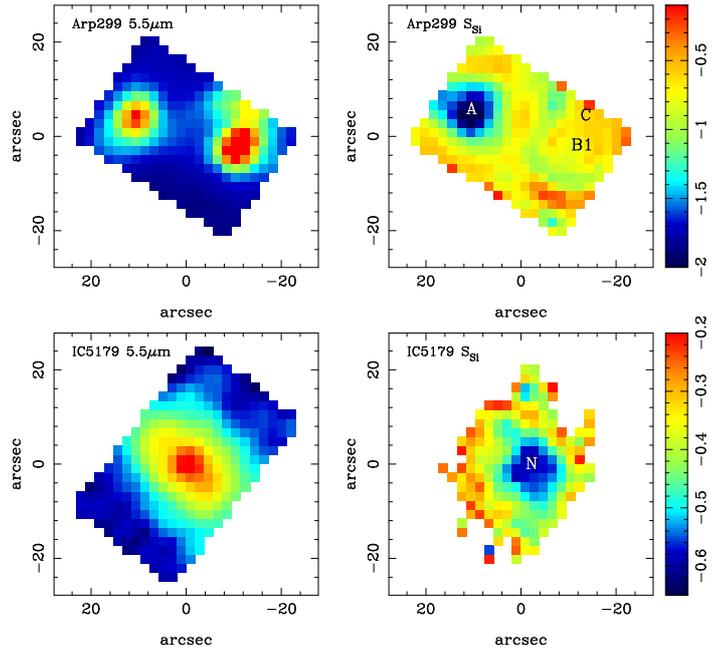}
\caption{IRS spectral maps of the continuum emission at $5.5\,\mu$m (left
  panels) and the apparent depth of the $9.7\,\mu$m silicate feature
  (right panels) of two LIRGs in our sample from the SL data cubes. 
For Arp~299 we mark the
  positions of the two nuclei: A for IC~694 (the eastern 
component) and B1 for NGC~3690 (the western component), as
  well as region C, an actively star forming region. For IC~5179 we
  mark the nuclear region as N. 
The maps have been rotated so the orientation is
  north up, east to the left.}
\end{figure}
\label{}

The majority of the observations were obtained in Cycle 3 and Cycle 4 as part
of a {\it Spitzer} guaranteed time observation (GTO) program (P.I.:
 G. H. Rieke, programs ID 30577 and ID 40479), 
except for the low spectral resolution observations (SL and LL
modules) of
 Arp~299, which were part of a different GTO program (P.I.: J. R. 
Houck, program ID 21) observed in  Cycle 1.

\begin{table}

\caption{IRS Observing Parameters}
 \begin{tabular}{lcccccccc}
\hline
Module & Step   & Perp  & Step
& Par  &
$t$ & Cycles  \\  
        & Perp  & Points  & Pars  &
        Points
  &        (s)       &   \\ 
(1) & (2) & (3) &(4) & (5) & (6) & (7) \\
\hline
SL1      &1.8  & 13 (17)     &  --  & --
& 14    & 2 \\
SL2      & 1.8  & 13 (17)     &  --  & --
& 14    & 2 \\ 
LL1      & 5.25 & 5  (7)      &  --  & --
& 14    & 2\\ 
LL2      & 5.25 & 5  (7)      &  --  & --
& 14    & 2\\ 
SH      & 2.35 & 9 (15)      &  5.65& 5
(7)       & 30    & 2 \\ 
LH      & 5.55 & 5  (7)      &   --   &
--        & 60    & 4\\ 
\hline
\end{tabular}

Notes.--- Column~(1): IRS Module. Column~(2): Size of the step along the
 direction perpendicular to the slit in arcseconds. Column~(3): 
Number of steps in the perpendicular direction for 
the $20\,{\rm arcsec} \times  20 \,{\rm
   arcsec}$ maps, and for the $30\,{\rm arcsec} \times  30\,{\rm
   arcsec}$ in brackets. Column~(4): Size of the step along the
 direction parallel to the slit in arcseconds. Column~(5): Number of steps in
 the parallel direction. Column~(6): Duration of the
 ramp in seconds. Column~(7): Number of cycles. 
\end{table}

We used the IRS spectral mapping capability, 
which involves moving the telescope 
perpendicular to the long axis of the slit
with a step of one-half the slit width until 
the  appropriate region is covered. 
Since in most cases the SL (length of 57 arcseconds) and LL 
(length of 168 arcseconds) slits are longer than the
extent of the galaxies,  we did not obtain separate 
background observations. For the LH module we obtained dedicated
background observations in staring mode (one single slit) at a
region about 2 arcminutes away from the galaxy. We did not observe 
backgrounds  for the SH module, except for those galaxies
  observed in Cycle 4 (NGC~6701, NGC~7591, and NGC~7771).
We mapped  approximately at least 
the central $20\,{\rm arcsec} \times 20\,{\rm arcsec}$ to
$30\,{\rm arcsec} \times 30\,{\rm arcsec}$ regions of the galaxies
(before rotation). 
The sizes of the maps were chosen to cover a large fraction of the mid-IR
emission of these galaxies. 
For the typical distances of our LIRGs, the IRS maps cover the 
central $\sim 3-11\,$kpc, depending on the galaxy and the IRS module. 
Table~2 summarizes the {\it Spitzer}/IRS 
observing parameters used for our GTO program.

We processed the data using the {\it Spitzer} IRS pipeline version
S15.3 for the SL and LL modules, and S17.2 for the SH and LH modules.
The data cubes were assembled using {\sc cubism}
(the CUbe Builder for IRS Spectra Maps, Smith et al. 2007b) from the
individual Basic Calibrated Data (BCD)  
spectral images. Full error cubes are also built alongside the data cubes 
by standard error propagation, using, for the input uncertainty, 
the BCD-level uncertainty estimates produced by the IRS pipeline  
from deviations of the fitted ramp slope fits for each pixel. These
uncertainties are used to provide error
estimates for the extracted spectra, and for the line and continuum maps (see
Smith et al. 2007b for full details). The angular resolutions (FWHM) of the
data cubes are $\sim 4$\,arcsec, and $\sim 5$\,arcsec for the SL and
 SH modules, respectively (see Pereira-Santaella et al. 2009a). For
the distances of our galaxies, these correspond to physical scales of
between $\sim 0.7\,$kpc for the nearest objects to $\sim 2\,$kpc for
the most distant ones.

\begin{figure}
\includegraphics[width=9cm]{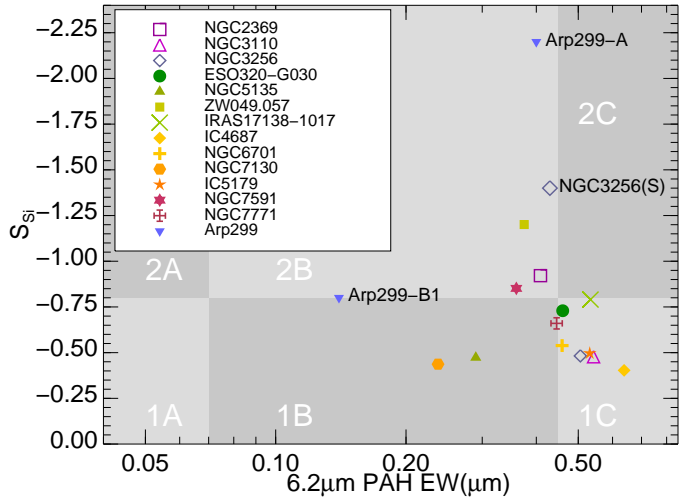}

\caption{Diagram of the EW of the $6.2\,\mu$m PAH feature 
versus the apparent depth of the $9.7\,\mu$m silicate feature 
for the nuclei of the galaxies in our sample. 
These quantities were  measured from  SL spectra extracted with 
$3.7\,{\rm arcsec} \times 3.7\,{\rm arcsec}$ apertures  centered
  at the nuclei of the galaxies.
The different regions in
this diagram are taken from Spoon et al. (2007). Region 1A is occupied
by AGN and it is characterized by low EW of the PAH feature and a weak
silicate feature. Region 1C has moderate  silicate feature in
absorption  to nearly-zero and high EW of the PAH, and it is mostly occupied by
starburst galaxies. Region 1B is intermediate between the previous
two, and it  is occupied by
galaxies with both AGN and star formation. Most ULIRGs in the Spoon et
al. (2007) sample fall in region 2B and region 3A. The 3A
(very deep silicate feature and low EW of the $6.2\,\mu$m PAH
feature), 3B and 3C regions of Spoon et al. (2007) are not
shown in this figure. }
\end{figure}

We extracted nuclear 1D spectra for all the IRS modules using 
{\sc cubism} with the smallest ($2 \times 2$ pixel) possible  
extraction apertures from the data cubes before rotation.
We also extracted  1D low-resolution (SL+LL) spectra 
covering approximately the extent of each system to be representative
of the mid-IR properties of the LIRGs (see Section~5). 

Full details on the observations, data reduction, and the 
construction of the data cubes, and spectral maps will be given by
Pereira-Santaella et al. (2009a).

\section{Characterizing the obscuration of LIRGs with the
  $9.7\,\mu$m  silicate feature}

LIRGs, ULIRGs and IR-bright galaxies are known to contain  highly obscured regions ($A_V \simeq
4-50\,$mag, see e.g., Veilleux et al. 1995; Genzel et al. 1998;
Alonso-Herrero et al. 2000,  2006a; Spoon et al. 2000, 2001; Roussel
et al. 2006; Sirocky et al. 2008), and are usually coincident with
the nucleus. In addition to the
"traditional" values derived using hydrogen recombination lines
(e.g., H$\alpha$/H$\beta$, Pa$\alpha$/H$\alpha$), we can derive the
distribution of the extinction using our spatially resolved maps of
the depth of the $9.7\,\mu$m absorption feature. The apparent 
strength of the $9.7\,\mu$m silicate feature is defined 
as $S_{\rm Si} = \ln f_{\rm obs} (9.7\mu{\rm m})/
f_{\rm cont}(9.7\mu{\rm m})$, where $f_{\rm cont}(9.7\mu{\rm m})$ is
the local continuum and $f_{\rm obs} (9.7\mu{\rm m})$ is the observed
flux density of
the feature. Negative values of the apparent strength
of the feature mean that the feature is observed in absorption,
whereas positive values mean that the feature is in emission. 
We used a method similar 
to that outlined by Spoon et al. (2007)  for
sources dominated by polycyclic aromatic hydrocarbon (PAH) 
emission, to fit the local continuum 
as a power law with pivot wavelengths at 5.5 and $14\,\mu$m.  
We extracted spectra  with  $2\,{\rm pixel} \times 2\,{\rm pixel}$
apertures, by  moving pixel by pixel along columns and rows until the
entire field of view was covered. We built the spectral maps of the silicate
feature by  measuring it on the individual 1D spectra. 
Assuming an extinction law and a dust geometry the silicate strength
can be converted into a visual extinction.
For comparison we  also constructed maps of the $5.5\,\mu$m continuum,
that is, of the mean flux of a bandpass 
covering the spectral range of $5.3-5.7\,\mu$m.

Figure~1 shows the spectral maps of the apparent strength of the
$9.7\,\mu$m silicate feature for two systems in our sample with very
different behaviors, together with the maps of the $5.5\,\mu$m
continuum. 
Arp~299 is the most luminous system in our sample
(Table~1). It is a pair interacting galaxies composed of IC~694 and
NGC~3690, whose nuclear regions are refered to as source A and source
B1, respectively. As can be seen from Fig.~1, this system shows 
a large range of values of the silicate feature. 
The region with the 
deepest silicate feature ($S_{\rm Si} \sim -2.2$, see 
Alonso-Herrero et al. 2009a for more details) is Arp~299-A  and has a value
comparable to the typical values of ULIRGs (see Armus et al. 2007;
Spoon et al. 2007, and also next section). The implied extinction
for Arp~299-A is 
$A_V\sim 36\,$mag for a foreground screen of dust (assuming
$A_V/\tau_{\rm Si} =16.6$; Rieke \& Lebofsky 1985), in agreement with
previous  works (see e.g., Gallais et al. 2004; Alonso-Herrero et
al. 2000). We note, however, that the 
measured silicate strength offers only a lower limit to
the true extinction (e.g., Levenson et al. 2007; Sirocky et
al. 2008). For Arp~299-B1, 
we measured $S_{\rm Si} \sim -0.8$, which is in the range
observed for Seyfert 2 galaxies (Shi et al. 2006; Hao et al. 2007).
The values of the apparent depth of the silicate feature 
throughout the rest of this system are relatively moderate and 
similar to the values measured in other starburst galaxies 
(Brandl et al. 2006; Spoon et al. 2007).

The other galaxy shown in Fig.~1 is IC~5179, a
spiral galaxy with a moderate IR luminosity (Table~1). The apparent
depth of the silicate feature in the nuclear region is 
$S_{\rm Si} \sim -0.5$, whereas in the circumnuclear regions the
silicate feature is shallower indicating a lower extinction. This
behavior is seen in a large fraction of the LIRGs in our sample (see
Pereira-Santaella et al. 2009a).

The nuclear values of the apparent depth of the silicate feature for
the galaxies in our sample can be seen in Fig.~2. For
the distances of our galaxies the SL nuclear extraction apertures probe
regions with physical sizes of between 0.6 and 1.4\,kpc, depending on
the galaxy. As can be seen from Fig.~2 the apparent depths of the
silicate features of the majority of the 
LIRG nuclei are moderate ($S_{\rm Si} \sim
-0.4 \, {\rm to}  \, -0.9$), and intermediate
between those observed in starburst galaxies (Brandl et al. 2006) and
ULIRGs (Spoon et al. 2007). Using the Rieke \& Lebofsky (1985)
extinction law and assuming a foreground screen of dust, the observed
apparent depths translate into  typical nuclear extinctions of
$A_V\sim 7-15\,{\rm mag}$.  For the most obscured nuclei in our sample:
Arp~299-A, the southern nucleus of NGC~3256, and Zw~049.057,
the nuclear extinctions are  in excess of $A_V\sim 20\,$mag (assuming
a foreground screen of dust).

\begin{figure}
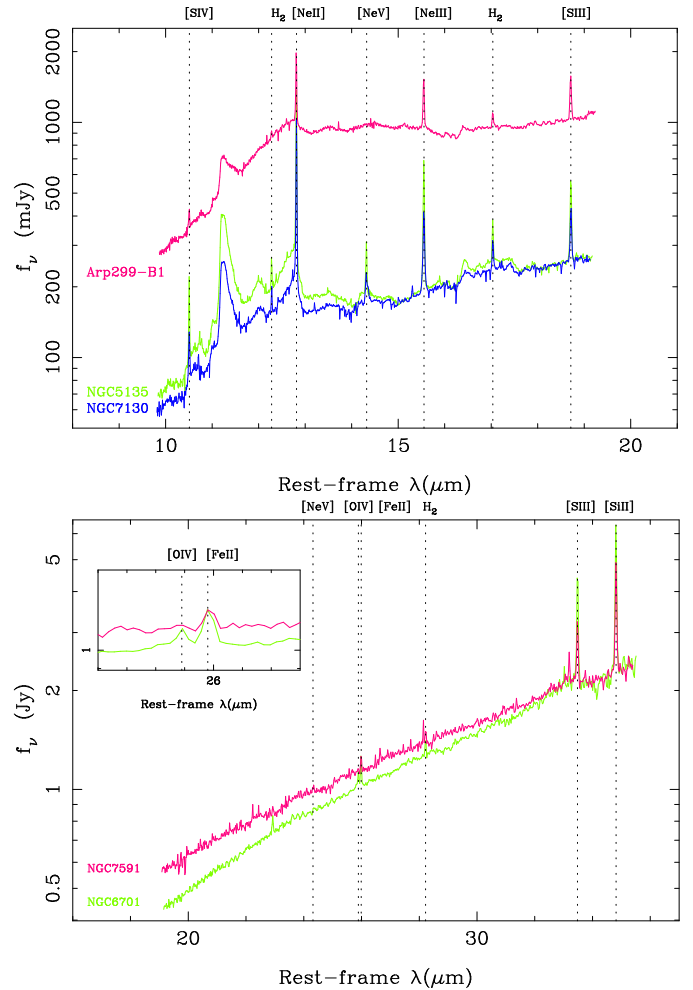

\includegraphics[width=6.5cm,angle=-90]{Fig3a.ps}
\includegraphics[width=6.5cm,angle=-90]{Fig3b.ps}

\caption{{\it Upper panel:} Rest-frame SH spectra of the three nuclei in our
  sample classified as  
Seyfert galaxies. The spectra were extracted with square apertures 
of $4.5\,{\rm arcsec }
\times 4.5\,{\rm arcsec}$ in size. The [Ne\,{\sc
  v}]$14.32\,\mu$m emission line is clearly detected in NGC~5135 and
NGC~7130, but not in Arp~299-B1. 
{\it Lower panel:} Rest-frame LH spectra of the two nuclei in our
  sample classified as composite (intermediate between   
LINER and H\,{\sc ii}) from optical spectroscopy (Alonso-Herrero et
al. 2009b). The spectra were extracted with square apertures 
of $8.9\,{\rm arcsec }
\times 8.9\,{\rm arcsec}$ in size. The [O\,{\sc
  iv}]$25.89\,\mu$m emission line is clearly detected in NGC~6701, but
it is very faint in the nuclear region of NGC~7591 (see inset).  
We also mark in the two panels the
positions of other bright mid-IR fine structure lines,  as well
  as molecular hydrogen lines.}

\end{figure}

\section{The AGN and Star Formation Properties of LIRGs}
\subsection{Identifying AGN in the mid-IR}
Composite (AGN + starburst) spectra are observed 
 in local LIRGs, and high-$z$ IR-bright galaxies (see e.g., Yan et al. 2005,
 2007; Dey et al. 2008). All our LIRGs except for
 NGC~2369 have a nuclear activity class derived from optical
 spectroscopy (see Table~1); three galaxies have a Seyfert
 classification, two are classified as composite, and the rest have
 an H\,{\sc ii}-like classification.  To compare with these results, 
there are a number of mid-IR diagnostics that can be used to test
for the presence of an 
AGN and to determine its contribution to the observed luminosity of the
system, as well as to identify highly obscured AGN.

AGN usually show high excitation lines. In the mid-IR the
brightest high excitation lines are  [Ne\,{\sc
  v}]$14.32\,\mu$m and  [O\,{\sc 
    iv}]$25.89\,\mu$m 
(e.g., Genzel et al. 1998;
Sturm et al. 2002; Mel\'endez et al. 2008). The large excitation 
potentials of the [Ne\,{\sc
  v}]$14.32\,\mu$m and the [O\,{\sc 
    iv}]$25.89\,\mu$m lines (97.1\,eV and 54.9\,eV, respectively, see
  Alexander et al. 1999)  make it unlikely that they are excited by
  star-formation. Moreover, the [O\,{\sc iv}]$25.89\,\mu$m line
  appears to be a good tracer of the AGN intrinsic luminosity 
(Mel\'endez et al. 2008; Diamond-Stanic et al. 2009; Rigby et al. 2009).
We note however that these high excitation emission lines are not always 
detected in relatively bright AGN (e.g., Weedman et al. 2005). 
IR-bright LINERs also show the [O\,{\sc 
    iv}]$25.89\,\mu$m  emission line, but their mid-IR spectral energy
  distributions are more similar to those of starburst galaxies (Sturm
  et al. 2006). Moreover, faint extended [O\,{\sc 
    iv}]$25.89\,\mu$m line emission has been detected in some starburst
galaxies, indicating that it can also be associated with star formation (Lutz
et al. 1998).

\begin{figure}

\includegraphics[width=8.7cm]{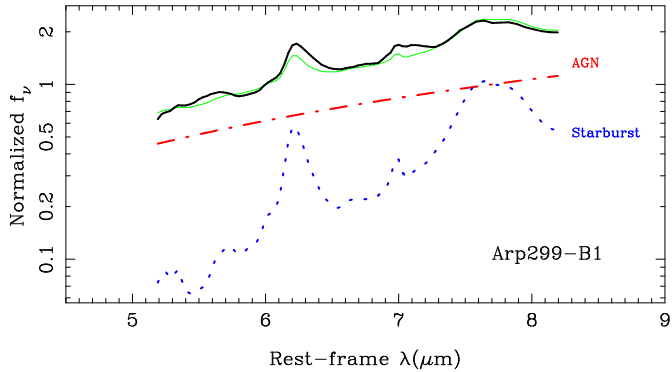}

\caption{Rest-frame $5-8\,\mu$m spectrum (thick black line)  
of Arp~299-B1 (see Fig.~1) 
observed with the SL1 module and 
extracted with a $3.7\,{\rm arcsec} \times 3.7\,{\rm arcsec}$ aperture.
The observed spectrum is normalized at $6\,\mu$m.
The observed spectrum is modeled (thin green line) 
as the sum of two components: 
an AGN continuum (dotted-dashed red line), and a 
starburst component (dotted blue line) using the Brandl et al. (2006)
template. This method is similar to 
that of Nardini et al. (2008). This figure has been adapted from a
similar figure presented by Alonso-Herrero et al. (2009a).}

\end{figure}

Figure~3 (upper panel) 
shows the SH nuclear spectra of the three nuclei in our sample
classified as Seyfert. The [Ne\,{\sc
  v}]$14.32\,\mu$m emission line is not 
detected in Arp~299-B1 (nuclear region of NGC~3690), and the upper limit is
consistent with that measured by Verma et al. (2003). In 
both NGC~5135 and NGC~7130 the [Ne\,{\sc
  v}]$14.32\,\mu$m line is clearly detected. 
This may be explained in terms of the lower
bolometric luminosity of the AGN in Arp~299-B1 when compared to the AGNs in
NGC~5135 and NGC~7130 (Levenson et al. 2004, 2005), as the [Ne\,{\sc
  v}]$14.32\,\mu$m appears to be a good indicator of the AGN
bolometric luminosity (Dasyra et al. 2008). 
The [Ne\,{\sc  v}]$14.32\,\mu$m line was not detected in any of the
two composite
nuclei. 

The [O\,{\sc 
    iv}]$25.89\,\mu$m line is detected in all the Seyfert nuclei in
  our sample. In the case of  composite nuclei (Fig.~3, lower panel), 
the [O\,{\sc  iv}]$25.89\,\mu$m emission line is very faint 
in NGC~7591, but it is clearly detected   
in NGC~6701. The latter galaxy contains a large number of bright H\,{\sc ii}
regions within the central 
($\sim 8.9\,{\rm arcsec} \times 
8.9\,{\rm arcsec}$, the LH extraction aperture) 
nuclear region (see Alonso-Herrero et 
  al. 2006a and 2009b), and the observed
[Ne\,{\sc ii}]$12.82\,\mu$m/[O\,{\sc iv}]$25.89\,\mu$m line
ratio  ($\sim 30$)
is  similar to, although slightly lower than, those observed in 
star-forming galaxies in our sample (Pereira-Santaella et
  al. 2009b). Thus, it is not clear 
what fraction, if any, of the observed  [O\,{\sc iv}]$25.89\,\mu$m
line emission is produced by the active nucleus in this galaxy. 
The possibility that faint [O\,{\sc 
    iv}]$25.89\,\mu$m line emission can also be produced by intense on-going
  star formation is further confirmed because this
  line is detected in all the nuclei in our sample 
classified as H\,{\sc ii} from optical
spectroscopy (see Pereira-Santaella et al. 2009b). Indeed, for the
H\,{\sc ii}-like nuclei in our sample of LIRGs, we
measured line ratios of  [Ne\,{\sc ii}]$12.82\,\mu$m/[O\,{\sc iv}]$25.89\,\mu$m
$>>10$, which are well explained by 
star formation  (see e.g., Genzel et al. 1998; Dale et al. 2006).

\begin{figure}
\includegraphics[width=8.6cm,angle=-90]{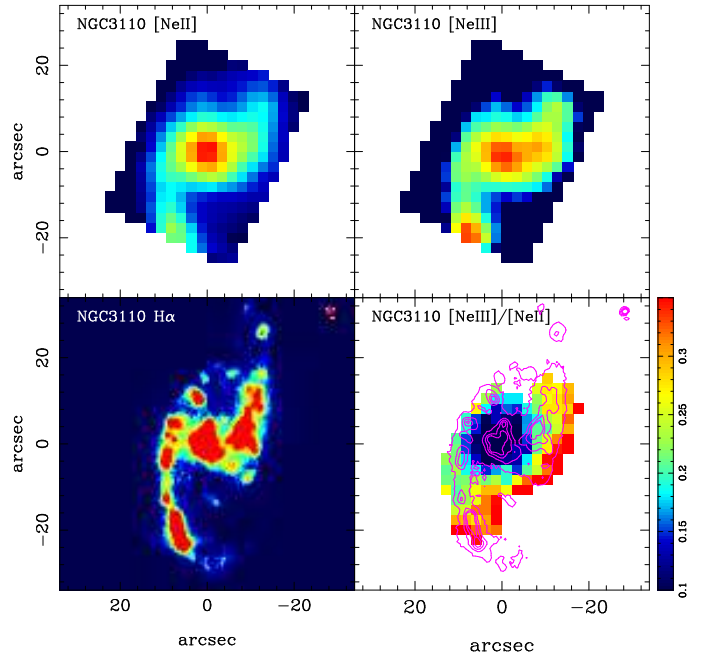}

\caption{The upper  panels are the observed 
IRS SH spectral maps of the [Ne\,{\sc
  ii}]$12.81\,\mu$m  (left) and [Ne\,{\sc iii}]$15.56\,\mu$m
(right) emission lines, both displayed in a square root scale.
For these two emission line maps, we only
    plot pixels with line peak  detections $\ge 1\sigma$ above the
    continuum, where $\sigma$ is the standard
  deviation of the continuum fit.
The lower left panel is the continuum-subtracted 
H$\alpha$ map of NGC~3110 from Hattori et al. (2004) shown in a square
root scale. The lower right panel is the SH observed (not corrected for
extinction) map of the [Ne\,{\sc iii}]$15.56\,\mu$m/[Ne\,{\sc ii}]$12.81\,\mu$m
line ratio, with the H$\alpha$ contours (in a linear
scale) superimposed.  Only pixels with line peak 
$\ge 3\sigma$  detections 
   in both emission lines are used to construct line ratio map. The
IRS maps have been rotated so the  
orientation is north up, east to the
left.}

\end{figure}

An alternative way to identify AGN is to use diagnostic diagrams such
as that of the equivalent width (EW) of the
 $6.2\,\mu$m PAH feature\footnote{We measured the EW of the
   $6.2\,\mu$m PAH feature by fitting the local continuum between 
$5.75$  and $6.70\,\mu$m and integrating the feature between 5.9 
and $6.5\,\mu$m. We calculated the EW by dividing the PAH flux
by the fitted local continuum at $6.2\,\mu$m.}
feature versus the apparent depth of the $9.7\,\mu$m 
silicate feature put forward by Spoon et al. (2007).  Figure~2 shows
such a diagram for the nuclear regions of our sample of LIRGs.  For
the distances of our galaxies the SL extraction apertures probe
regions with physical sizes of between 0.6 and 1.4\,kpc, depending on
the galaxy. The three Seyfert nuclei in our sample 
(Arp~299-B1, NGC~5135, and NGC~7130)
are located in the
2B region of this diagram, between the region occupied by AGN and that
occupied by starburst galaxies, that is, the regions 1A and 1C,
respectively (see Spoon et al. 2007 for details). 
This is in agreement with previous results showing the
composite (AGN + star formation) 
nature of these nuclear regions  (see Gonz\'alez-Delgado et al. 1998; 
Garc\'{\i}a-Mar\'{\i}n et al. 2006;
Alonso-Herrero et al. 2006a,b, 2009a; Bedregal et al. 2009  
and references therein). Note also the
$11.3\,\mu$m PAH feature in their nuclear spectra (see Fig.~3, upper panel),
clearly indicating the presence of star formation. 

The majority of the 
nuclei in our sample classified as H\,{\sc ii} or composite 
are located in the 1C starburst region, while only 
a few nuclear regions (i.e., NGC~2369, Arp~299-A,
NGC~3256(S), Zw049.057, and NGC~7591) lie in one
of the regions occupied by ULIRGs (the 2B
region, see Fig.~2).
 There are no examples in our sample of LIRG nuclei with very
deep silicate features ($S_{\rm Si}< -2.5$) and low EW ($<0.07\,\mu$m)
of the $6.2\,\mu$m PAH feature, that is, the 3A
region also occupied by ULIRGs (Spoon et al. 2007).

We can use a diagnostic method similar to that of Nardini et
al. (2008) to quantify the relative strengths of AGN and starbursts in
nuclei with both. This method is based on the close similarity of the
rest-frame 
$5 - 8\,\mu$m spectra of the high (approximately solar or
  supersolar, see Bernard-Salas et al. 2009) 
metallicity low-$z$ starbursts observed by Brandl et
al. (2006). Thus, any excess in the
$5-8\,\mu$m spectral region can be attributed to continuum emission from hot
dust. Nardini et al. (2008) interpreted this  hot dust component as an 
indication for the presence of an AGN.  For our 
estimate we used two different templates, the  starburst template 
of Brandl et
al. (2006) and the star-forming ULIRG template of Nardini et
al. (2008). Figure~4 shows an example of this method applied to source B1 in
Arp~299, the nuclear region of NGC~3690. While we did not detect
the high excitation [Ne\,{\sc v}]$14.32\,\mu$m line in this nucleus, the 
AGN is clearly identified via the presence of a strong dust
component. This strong dust component is even detected at shorter
wavelengths $2-3\,\mu$m (see Alonso-Herrero et al. 2000; Soifer et
al. 2001). We estimated that the AGN in Arp~299-B1
accounts for $\sim 80-90\%$ of the observed  
emission at $6\,\mu$m (see Alonso-Herrero et
al. 2009a) within a $3.7\,{\rm arcsec}  \times 3.7\,{\rm arcsec}$  aperture.

\begin{figure}
\includegraphics[width=5.cm,angle=-90]{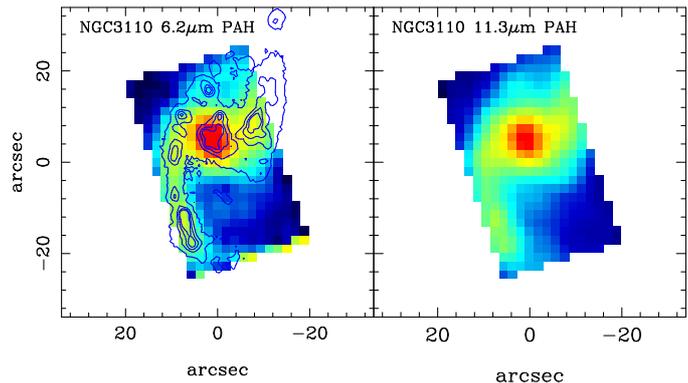}

\caption{IRS SL observed (not corrected for extinction) 
spectral maps of the $6.2\,\mu$m (left) and the
  $11.3\,\mu$m PAH features in NGC~3110, shown in a square root
  scale. The contours are the
  H$\alpha$ emission as in Fig.~5. For both maps, we only
    plot pixels above $1\sigma$. The IRS maps have been rotated so
  the orientation is north up, east
  to the left.}

\end{figure}

\subsection{Star formation properties}

\subsubsection{Fine structure emission lines}
Mid-IR fine structure emission lines can be used to probe the star
formation processes in galaxies, with the added advantage that they 
are less affected by extinction
than optical (e.g., H$\alpha$) and near-IR (e.g., Pa$\alpha$, 
Br$\gamma$) emission lines. In particular, the 
[Ne\,{\sc ii}]$12.81\,\mu$m  line (Roche et al. 1991; 
  D\'{\i}az-Santos et al. 2009), or  
alternatively 
the sum of the [Ne\,{\sc ii}]$12.81\,\mu$m and the [Ne\,{\sc
    iii}]$15.56\,\mu$m emission lines, for very young stellar populations, 
 appear to be good tracers of the ionizing stars (Ho \& Keto 2007 and references
therein). Moreover, because the  
[Ne\,{\sc iii}]$15.56\,\mu$m/[Ne\,{\sc ii}]$12.81\,\mu$m 
line ratio is sensitive to the
hardness of the radiation field, it can be used as an indicator of the
age of the stellar population.
We note however, that this ratio also
depends on  the metallicity, nebular density and
ionization parameter (see e.g., Thornley et al. 2000;
Mart\'{\i}n-Hern\'andez et al 2002; Rigby \& Rieke
2004; Snijders et al. 2007). 

\begin{figure*}
\hspace{2cm}
\includegraphics[width=7.cm,angle=-90]{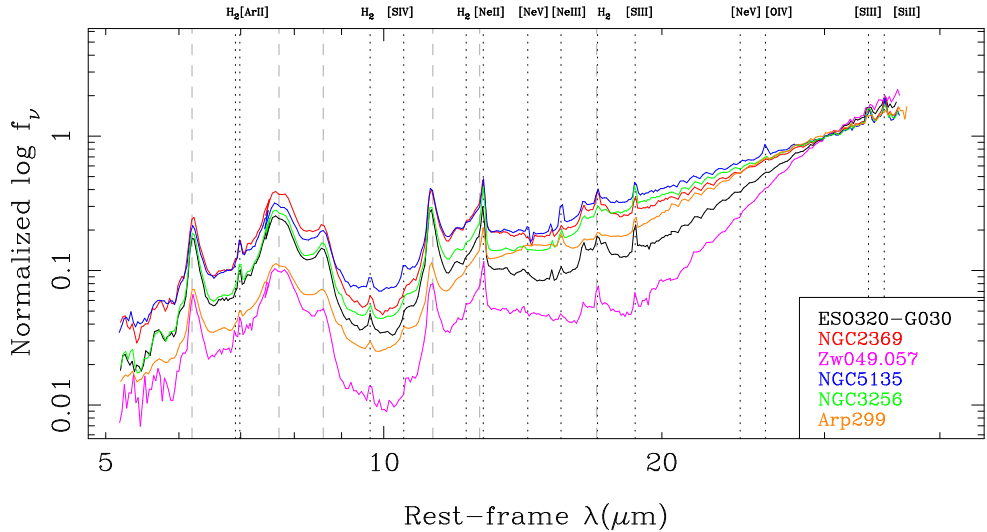}

\caption{Integrated low resolution (matched SL+LL) 1D spectra of six LIRGs in
  our sample spanning the full range of IR luminosity and normalized
  at $30\,\mu$m. We mark the
  positions of the brightest mid-IR emission lines (dotted lines), as
  well as the $6.2\,\mu$m and 8.6$\,\mu$m PAH features and 
the $7.7\,\mu$m, $11.3\,\mu$m, $12.7\,\mu$m, 
and $17\,\mu$m PAH complexes (dashed lines).}

\end{figure*}

We find that in general the brightest mid-IR emission lines, 
[Ne\,{\sc ii}]$12.81\,\mu$m, [Ne\,{\sc iii}]$15.56\,\mu$m, and
[S\,{\sc iii}]$18.71\,\mu$m,  have an overall morphology similar to that of H$\alpha$ and
Pa$\alpha$ (see Alonso-Herrero et al. 2009a and Pereira-Santaella et
al. 2009a). Figure~5 shows an example for one of the LIRGs in our
sample: NGC~3110. There is  [Ne\,{\sc ii}]$12.81\,\mu$m  line emission
arising from the nuclear region as well as in H\,{\sc ii} regions as traced by
regions of high surface brightness H$\alpha$ emission in the spiral
arms. As found by Alonso-Herrero et al. (2009a) for Arp~299, the [Ne\,{\sc
  iii}]$15.56\,\mu$m line emission of NGC~3110 seems to have a better
morphological correspondence with the hydrogen recombination line
emission (e.g., the H\,{\sc ii} regions west and south of the nucleus).
Another fundamental difference is that the [Ne\,{\sc ii}]$12.81\,\mu$m
emission line  is dominated by the nuclear emission,
while the H$\alpha$ map shows that the nucleus and 
extra-nuclear regions have a similar brightnesses.  This 
indicates that the H$\alpha$ nuclear emission is suffering from
significant extinction in the optical. The observed apparent depth ($S_{\rm
  Si}=-0.45$, see Fig.~2) of the $9.7\,\mu$m
feature implies $A_V\sim 7\,$mag, while the extinction derived from
optical and near-IR hydrogen recombination lines is  $A_V\sim 3\,$mag (Veilleux
et al. 1995; Alonso-Herrero et al. 2006a).

We also show in Fig.~5  the SH map of the observed
[Ne\,{\sc iii}]$15.56\,\mu$m/[Ne\,{\sc ii}]$12.81\,\mu$m line ratio
for NGC~3110. This ratio ranges from 0.1 to approximately
0.4 for NGC~3110, which is similar to the range observed 
in the majority of the star-forming LIRGs ($\sim 0.1-0.8$) in our sample (see
Pereira-Santaella et al. 2009a) and other high-metallicity 
starburst galaxies (see Thornley et al. 2000; Verma et al. 2003;
Brandl et al. 2006). In our non-AGN
LIRGs, the lowest values of this ratio 
are generally coincident with the nuclear regions,
whereas the largest values are associated with extra-nuclear H\,{\sc
  ii} regions and regions of diffuse low surface brightness  
H$\alpha$ or Pa$\alpha$ emission (see Fig.~5, and also Alonso-Herrero
et al. 2009a). The relatively low values of this ratio in most of the
nuclear regions of LIRGs could be explained in terms of 
density effects, as for a given age the ratio decreases for increasing
densities; the effect is most noticeable at high
densities\footnote{
The critical densities are  $1.8\times 10^5\,{\rm 
  cm}^{-3}$ and $6.1\times 10^5\,{\rm  cm}^{-3}$ for [Ne\,{\sc
  iii}]$15.56\,\mu$m 
and [Ne\,{\sc ii}]$12.81\,\mu$m, respectively.} 
($n_{\rm H}>10^4\,{\rm cm}^{-3}$,  
see Snijders et al. 2007). Other effects include 
the presence of relatively evolved starbursts
(see discussions by Thornley et al. 2000 and 
Alonso-Herrero et al. 2009a) and metallicity effects (see Verma et
al. 2003; Snijders et
al. 2007). Alternatively, the youngest stars may still be 
embedded in ultra-compact H\,{\sc ii} regions where the extinction is
very high and the
 ionized gas is at densities exceeding the critical densities
 of the neon lines (see Rigby \& Rieke 2004). 

Higher 
[Ne\,{\sc iii}]$15.56\,\mu$m/[Ne\,{\sc ii}]$12.81\,\mu$m line ratios
are seen at 
increasing galactocentric distances and in regions of diffuse ionized
gas in LIRGs (see Alonso-Herrero et al. 2009a for Arp~299), 
and also in other galaxies:  NGC~253 (Devost et al. 2004),   
M82 (Beir\~ao et al. 2008), and NGC~891 (Rand et
al. 2008). These high ratios
may also be related to the mechanisms
responsible for the increased optical line ratios 
(e.g., [S\,{\sc ii}]$\lambda\lambda$6716,6731/H$\alpha$, 
see a review by Haffner et al. 2009) in regions of diffuse
emission, although current
models are not able to reproduce the observed 
[Ne\,{\sc iii}]$15.56\,\mu$m/[Ne\,{\sc ii}]$12.81\,\mu$m line ratios
(see Binette et al. 2009).

\subsubsection{PAH feature emission}

The mid-IR spectral range  contains a large number of PAH
features, with the most prominent ones  being at 6.2, 7.7, 8.6 and
$11.3\,\mu$m. High metallicity star forming galaxies are
known to show bright PAH features, while the PAH emission is faint in
low-metallicity galaxies (see e.g., Roche et al. 1991; 
Madden et al. 2006; Engelbracht et
al. 2005; Smith et al. 2007a; Wu et al. 2006; Calzetti et al. 2007). 
The good
correlation between the integrated PAH emission and the IR luminosity of high
metallicity star-forming galaxies   and ULIRGs indicates that the PAH
emission is tracing the star formation activity 
at least on large scales (Brandl et al. 2006; Smith et al. 2007a;
Farrah et al. 2007). 

All the LIRGs in our sample show bright
$6.2\,\mu$m and $11.3\,\mu$m PAH emission over the field of view
mapped with the IRS. The overall morphologies of the PAH maps
are similar to those of the mid-IR fine structure lines and hydrogen
recombination lines (see Figs.~5 and 6),  although we are limited by the
physical scales of a few kpc probed by the IRS spectral mapping
observations. 
This indicates that, at least to first order, the PAH features are tracing
the star formation activity of LIRGs. Indeed, Peeters et al. (2004)
showed that although PAHs may trace B stars better than O stars, the
integrated PAH emission of galaxies 
appears to be a good overall tracer of the star
formation rate. On 
smaller scales (tens-hundreds of parsecs), however, 
PAH emission can stem from sites with no evidence for on-going massive star
formation (Tacconi-Garman et al. 2005; D\'{\i}az-Santos et
al. 2009).

Theoretical models predict that the relative strengths of the different PAH
features depend on the properties of the dust grains and 
the charging conditions, as well as the starlight 
intensity (Draine \& Li 2001). In particular, neutral PAHs are
expected to show large  $11.3\,\mu$m to $6.2\,\mu$m ratios, whereas ionized PAHs have smaller ratios. 
The spatially resolved maps of the $11.3\,\mu$m/$6.2\,\mu$m PAH ratio show 
variations within galaxies on scales of a few kiloparsecs. 
In some galaxies these variations in the observed PAH ratios 
can be readily attributed to the effects of the extinction (see
also Rigopoulou et al. 1999; 
Brandl et al. 2006; Beir\~ao et al. 2008), as is the case for 
Arp~299-A (see Alonso-Herrero et al. 2009a and Fig.~1) and
other galaxies in the sample. We find however, that in other LIRGs
there is no correlation between the observed
$11.3\,\mu$m/$6.2\,\mu$m PAH ratio 
and the apparent depth of the $9.7\,\mu$m silicate feature,
and it is possible that the variations of this ratio may be showing
variations in the  physical conditions within galaxies, and from
galaxy to galaxy. In a forthcoming paper, Pereira-Santaella et
al. (2009a), we will discuss fully this subject for our sample of LIRGs.

\section{Integrated mid-IR spectra of LIRGs}

Since the IRS maps cover all or nearly all of the mid-IR emitting
regions in these 
galaxies, we combined the full SL+LL data cubes into single mid-IR
spectra for each galaxy. We show a few examples covering the full
range of IR luminosities of the integrated
mid-IR spectra in our sample in Fig.~7, as an illustration of the
variety of spectral shapes. The spectra have been normalized at
$30\,\mu$m for an easy comparison with the typical mid-IR spectra of
ULIRGs presented by Armus et al. (2007, their figure~2, lower panel). 
It is clear from Fig.~7 that the integrated spectra of the 
majority of LIRGs do not show the deep silicate features characteristic
of ULIRGs (see also Fig.~2), and are quite similar to the mid-IR
spectra of local starburst galaxies (Fig.~8). 
It is also apparent that there is
no clear relation between the IR luminosity and  the
apparent depth of the silicate feature. For instance, the integrated 
spectrum of Zw~049.57 shows the
deepest silicate feature, whereas  the most luminous system in our sample,
Arp~299, has an apparent depth of the silicate feature of $S_{\rm Si}
\sim -0.8$ (compare with the apparent depth of Arp~299-A, see Fig.~1). 

In what follows, we describe a few examples of
the potential applications of these kinds of templates.

\subsection{Comparison with IR-bright galaxies at high-$z$}
One of the most surprising results obtained with very sensitive {\it
  IRS} observations is that a large fraction of  
IR-selected galaxies at high-$z$ are classified
as ULIRGs in terms of their IR luminosities,
but their mid-IR spectra are more similar to those of local
star-forming galaxies (e.g., Yan et al. 2005, 2007; Farrah et
al. 2008; Rigby et al. 2008). Specifically, high-$z$ ULIRGs show
strong PAH features and moderate depths of
the $9.7\,\mu$m silicate feature, while their local counterparts 
tend to show very deep silicate features and moderate to 
low equivalent widths of
the PAHs (e.g., Spoon et al. 2007; Armus et al. 2007; 
Farrah et al. 2007). 
\begin{figure}
\includegraphics[width=6.3cm,angle=-90]{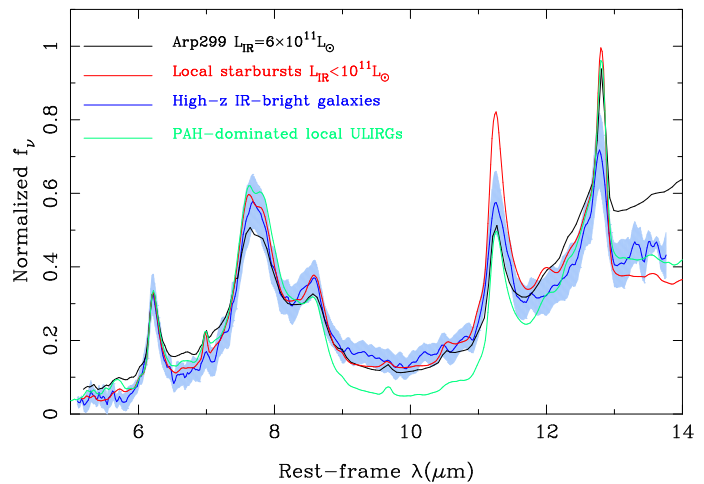}

\caption{Comparison of the low spectral-resolution SL 
spectra of the integrated
  emission of Arp~299 (black line) with the average   
template of local starbursts (red line) of Brandl et 
  al. (2006), the average spectrum of PAH-dominated local ULIRGs (the
  2C class of Spoon et al. 2007, see also Fig.~2,  green line), 
and the average spectrum and corresponding 
$1\sigma$ dispersion of high-$z$ IR-bright galaxies 
(blue line and shaded  blue area) 
from the Farrah et al. (2008) sample. The IR ($1-1000\,\mu$m) 
luminosity range of the 
galaxies included in this high-$z$ template is $\sim 8 \times 10^{12}$ to 
$6 \times 10^{13}\,{\rm L}_\odot$. The
  spectra have been scaled to match the peak of the $6.2\,\mu$m PAH
  feature. This figure has been adapted from Alonso-Herrero et al. (2009a).}

\end{figure}

In Alonso-Herrero et al. (2009a) we showed that 
the integrated spectrum of one of the galaxies in our sample: Arp~299, 
and the average spectrum of  the Farrah et
al. (2008) high-$z$ ULIRGs are remarkably similar (see Fig.~8). 
One of the favored explanations for the observed mid-IR spectra of high-$z$
ULIRGs is that star formation is diffuse and extended,  
or distributed in multiple dusty star-forming regions spread over
 several kiloparsecs (Farrah et al. 2008; Rigby et al. 2008). A
 similar behavior is found for the mid-IR spectra of submillimeter
 galaxies, which are again interpreted as arising from regions with
 sizes of a few kiloparsecs  (see
 Men\'endez-Delmestre et al. 2009). 
This is reminiscent of Arp~299, where the star formation is spread 
across at least 6-8\,kpc (see Soifer et al. 2001; 
Gallais et al. 2004; Alonso-Herrero et al. 2009a), 
with a large fraction taking place 
in regions of moderate mid-IR optical
depths (Figure~1, upper panel). 
The majority of the local LIRGs in the
 representative sample of Alonso-Herrero et al. (2006a)  behave 
in this manner, that is, the mid-IR emission is more extended 
(a few kpc, Alonso-Herrero et al. 2006b; D\'{\i}az-Santos et al. 2008) than the
highly embedded nuclear regions on 
sub-kiloparsec scales typical of local ULIRGs (Soifer et al. 2000). 
In Arp~299, it is only the nuclear region of IC~694 (Arp~299-A) 
that shows some of the typical mid-IR
characteristics of ULIRGs (see Figure~2), 
that is, very compact and embedded star
formation resulting in a deep silicate feature and
moderate EW of the PAHs. Arp~299 then may be a 
local counterpart of the star formation processes taking place in 
high-$z$ IR-bright galaxies (see also
 Charmandaris,  Le Floc´h, \& Mirabel 2004), albeit with a lower IR
 luminosity and possibly higher metallicity.

\subsection{Star Formation Rates for IR galaxies}

Observations at $24\,\mu$m with the Multiband Imaging Photometer for {\it
  Spitzer}  (MIPS, Rieke et al. 2004) 
have proven to be extremely successful in identifying
high-$z$ IR bright galaxies. One of the main problems faced by
these high-$z$ studies is to determine accurate SFRs
for dusty galaxies where UV and optical SFR indicators can be
severely affected by extinction effects. Moreover, because of the
variety of spectral energy distributions (SEDs) of these IR bright galaxies,
and the current lack of sensitive observations at $\lambda > 24\,\mu$m, one
has to make assumptions about the IR SEDs (e.g., use theoretical templates) 
to convert the observed monochromatic IR luminosities into
total IR luminosities.

\begin{figure}
\includegraphics[width=8.7cm]{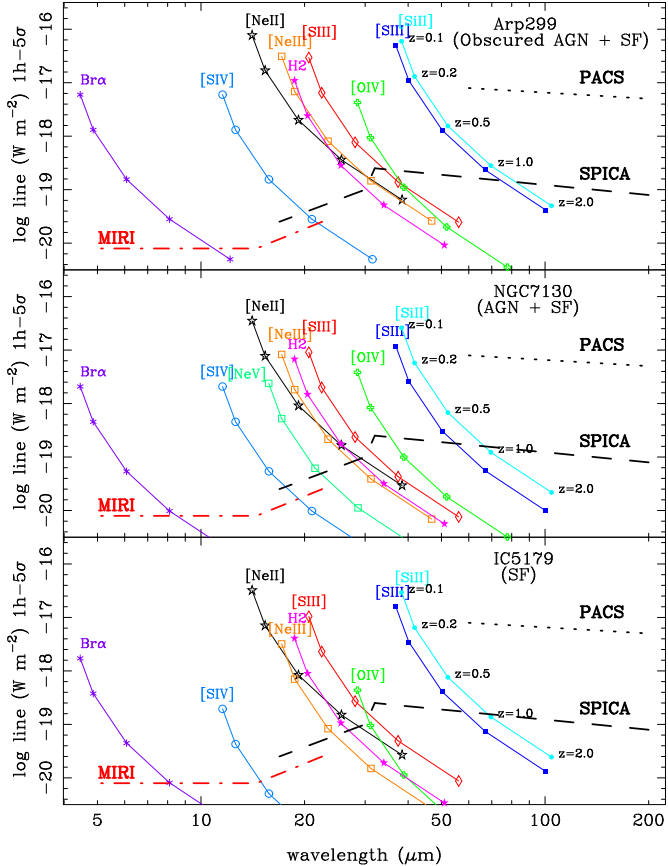}

\caption{Predictions as a function of redshift 
of the detectability ($5\sigma$) of Br$\alpha$, 
bright mid-IR fine structure
  lines and the brightest molecular hydrogen line (H$_2$ S(1)) 
in LIRGs for future IR missions for 1 hour integration
  time. 
The predicted sensitivity curves are from Wright et al. (2008) for 
{\it JWST}/MIRI (dotted-dashed line), from Swinyard et al. (2009) for
SPICA (dashed line) and from Poglitsch et al. (2008) 
for {\it Herschel}/PACS (dotted
line). We simulated the following emission lines: 
Br$\alpha$ at $4.05\,\mu$m, [S\,{\sc iv}]$10.51\,\mu$m, [Ne\,{\sc
  ii}]$12.81\,\mu$m, [Ne\,{\sc v}]$14.32\,\mu$m, [Ne\,{\sc
  iii}]$15.56\,\mu$m, H$_2$ S(1) at $17\,\mu$m, [S\,{\sc
  iii]}$18.71\,\mu$m,  [O\,{\sc iv}]$25.89\,\mu$m, 
[S\,{\sc iii}]$33.48\,\mu$m, and [Si\,{\sc ii}]$34.82\,\mu$m. We assumed
$H_0=71\,{\rm km\,s}^{-1}\,{\rm
  Mpc}^{-1}$, $\Omega_{\rm M}=0.27$, and $\Omega_\Lambda=0.73$ for
these simulations. 
The simulations are for the integrated line fluxes of three
  LIRGs in our sample: Arp~299 with $L_{\rm IR} \sim 8 \times 10^{\rm 11}\,{\rm
    L}_\odot$, NGC~7130 with $L_{\rm IR} \sim 3 \times 10^{\rm 11}\,{\rm
    L}_\odot$, and IC~5179 with $L_{\rm IR} \sim 2 \times 10^{\rm 11}\,{\rm
    L}_\odot$. }

\end{figure}

In a recent paper, Rieke et al. (2009), we used
the observed mid-IR spectra of some of the local LIRGs discussed here,
as well as of starburst galaxies (from Brandl et al. 2006) and ULIRGs
(from Rigopoulou et al. 1999 and Armus et al. 2007), 
together with UV, optical and
near-IR photometry to construct average templates covering the range in IR
luminosities of $9.75 \le \log L_{\rm IR} ({\rm L_\odot}) \le 13$. The main
criteria for the selected galaxies were that: (1) they have good high quality
data covering the full electromagnetic spectrum, from the UV to
submillimeter wavelengths, and (2) they are only powered by star
formation. These
templates were then used to determine the relation between the
observed mid-IR flux density and the SFR as a function of
the redshift of the galaxy.  These predictions were computed for a
number of 
  different mid and far-IR filters,  with wavelengths between 18 and
  $100\,\mu$m, of MIPS on {\it Spitzer}, PACS (Poglitsch et al. 2008) on {\it
    Herschel}, and MIRI (Wright et al. 2004) 
on the James Webb Space Telescope ({\it
    JWST}). We refer the reader to Rieke et al. (2009) for full details.

\subsection{Prospects for future space IR missions}
Another interesting application of the integrated mid-IR spectra of
local LIRGs is to make predictions for future mid and far-IR missions
of the detectability of bright mid-IR fine structure emission 
lines for high-$z$ LIRGs,
again to take advantage of the reduced extinction for these dusty
objects. This is of particular interest because a large fraction of 
high-$z$ IR-selected galaxies, submillimeter galaxies and massive galaxies 
as well as active galaxies appear to be powered by  
both AGN and star formation activity (e.g., Polletta et al. 2006; 
Yan et al. 2007; Daddi et al. 2007; Pope et 
al. 2008a,b; Dey et al. 2008; Alonso-Herrero et al. 2008; Ramos
Almeida et al. 2009). Since so far these results have been  
mostly based on the observed SEDs and/or low
spectral resolution IRS spectroscopy, disentangling the AGN and star formation
contributions can be challenging.  Future IR missions will have the
sensitivity and spectral resolution to detect mid-IR emission lines
from high-$z$ galaxies. 
As an illustration, we have simulated their detectability for three IR
instruments using the current predicted sensitivities: 
PACS (Poglitsch et al. 2008) on {\it Herschel}, 
MIRI (Gillian Wright, 2009, private communication) on the {\it
  JWST}, and SPICA (Swinyard et al. 2009).

We have chosen three LIRGs in our sample to cover a range of IR
luminosities as well as different mechanisms contributing to the IR
emission. Arp~299 has an IR luminosity close to the ULIRG limit and 
contains regions of intense star formation as well
as at least one low-luminosity ($L_{\rm 2-10keV} = 1.1 \times 
10^{41}\,{\rm erg \, s}^{-1}$, Ballo et al. 2004) obscured
AGN. NGC~7130 has an intermediate IR 
luminosity and  hosts a Compton-thick AGN (observed luminosity 
$L_{\rm 2-10keV} = 8.9  \times 
10^{40}\,{\rm erg \, s}^{-1}$ and estimated intrinsic AGN luminosity 
$L_{\rm 2-10keV} \sim 
10^{43}\,{\rm erg \, s}^{-1}$, Levenson et al. 2005) and star formation in the
nuclear region and spread throughout the galaxy. 
IC~5179 is a moderately luminous IR system and appears to be only
powered by star formation. For all these three galaxies we measured the 
fluxes of the brightest fine structure emission lines from the
integrated SH and LH spectra (see Pereira-Santaella et al. 2009a,b). 
We computed the Br$\alpha$ fluxes from
the extinction-corrected Pa$\alpha$ fluxes reported
  by Alonso-Herrero et  al. (2006a), 
 assuming case B recombination. These fluxes were measured 
over  regions  with sizes similar to those mapped with the IRS.

In Fig.~9 we present the results for the detectability ($5\sigma$, 1
hour  integration time) of the
brightest mid-IR emission lines of LIRGs as a function of redshift and the
three IR instruments. The simulations assumed that the line emission
is unresolved, and thus, for the majority of the lines the
detectability is an upper limit as the line emission is expected to be
extended in most cases. The tracers of star formation  (i.e., 
Br$\alpha$, [Ne\,{\sc ii}]$12.81\,\mu$m and [Ne\,{\sc
  iii}]$15.56\,\mu$m) in LIRGs will be detected with MIRI and SPICA 
 out to redshift of $z\sim 1$. These lines have  the advantage of both
 a reduced extinction  and a relatively straightforward 
interpretation in terms of the current SFR. 
The [Ne\,{\sc v}]$14.31\,\mu$m line, which provides definite
evidence for the presence of an AGN, will be detected out to $z\sim
0.5$ with MIRI and SPICA 
for Compton-thick AGN hosted by LIRGs and other types galaxies (e.g.,
X-ray selected AGN).  Finally other bright mid-IR lines such as the
[S\,{\sc iii}]$18.71\,\mu$m, [S\,{\sc iii}]$33.48\,\mu$m, and 
[Si\,{\sc ii}]$34.82\,\mu$m lines 
and the $17\,\mu$m H$_2$ S(1) transition. The
latter is a good tracer of the warm molecular gas in the ISM,  and will be
easily detected out to at least $z\sim 0.5$ with SPICA.

\section{Conclusions}
We  presented a GTO {\it Spitzer}/IRS
program aimed to obtain mid-IR ($5-38\,\mu$m) spectral mapping of a sample of
14 local LIRGs covering at least the central $20\,{\rm arcsec} \times 20\,{\rm arcsec}$ 
to $30\,{\rm arcsec} \times 30\,{\rm arcsec}$  regions. 
We used all four IRS modules: SL, LL, SH, and LH.
We constructed spectral maps of the bright mid-IR emission
lines (e.g., [Ne\,{\sc ii}]$12.81\,\mu$m, [Ne\,{\sc iii}]$15.56\,\mu$m, 
[S\,{\sc iii}]$18.71\,\mu$m, H$_2$ at $17\,\mu$m),  
continuum, the 6.2 and $11.3\,\mu$m 
PAH features, and the $9.7\,\mu$m silicate feature. 
The physical scales resolved 
by the short-wavelength module maps (SL and SH) are
between $\sim 0.7\,$kpc for the nearest objects to $\sim 2\,$kpc for
the most distant LIRGs in our sample. We also extracted 
1D spectra of regions of interest in each galaxy. 
The main goal of this paper is to describe the goals and present the
first results of this  {\it Spitzer}/IRS  GTO program.

For a large fraction of the LIRGs in our sample, the regions with the
deepest $9.7\,\mu$m silicate feature are coincident with the nuclei of
the galaxies.  The nuclear values of the apparent depth of the
silicate feature are moderate (mostly between $S_{\rm Si} \sim -0.4$ to $S_{\rm
  Si}\sim -0.9$), and thus are
intermediate between the heavily absorbed nuclei of ULIRGs and the
relatively shallow silicate features of starburst galaxies.  Only
three nuclei, 
Zw~049.057, Arp~299-A  and NGC~3256(S), have deep silicate
features ($S_{\rm Si} <-1$). 

Since all but one nuclei have a nuclear activity classification from
optical spectroscopy, we 
applied a number of diagnostics to test for the  
presence of an AGN. In particular we looked for  the high excitation
 [Ne\,{\sc v}]$14.32\,\mu$m and [O\,{\sc
  iv}]$25.89\,\mu$m emission lines and the presence 
of a strong dust continuum emission at $6\,\mu$m. We found that
for the three nuclei classified as Seyfert the 
mid-IR diagnostics gave a consistent classification. 
We detected [O\,{\sc iv}]$25.89\,\mu$m emission in most of the LIRG
nuclei classified as H\,{\sc ii}-like or composite (intermediate
  between LINER and H\,{\sc ii}) 
from optical spectroscopy, as well as in the Seyfert nuclei. We
determined, however, that the large observed [Ne\,{\sc
  ii}]$12.81\,\mu$m/[O\,{\sc iv}]$25.89\,\mu$m line ratios ($>>10$)
of H\,{\sc ii}-like nuclei are consistent with being 
produced by star formation processes rather than by an AGN. 

The morphologies of  [Ne\,{\sc ii}]$12.81\,\mu$m, [Ne\,{\sc
  iii}]$15.56\,\mu$m  
and the PAH features (at 6.2 and $11.3\,\mu$m) of LIRGs are 
similar to those of hydrogen recombination lines, on
the scales probed by the IRS data --- a few kiloparsecs. This indicates
that the integrated [Ne\,{\sc ii}]+[Ne\,{\sc iii}] emission, as well
as the PAH emission trace well the star formation activity in this
kind of galaxies. 

We finally used the integrated {\it Spitzer}/IRS spectra 
as templates of local LIRGs. We showed that the 
integrated spectra of local LIRGs and in particular that of Arp~299, are
similar to those of high-$z$ ULIRGs. This adds support to the
scenario where star-formation in high-$z$ IR-bright galaxies is spread
over several kiloparsecs, rather than in the highly embedded compact
($<1\,$kpc) nuclear regions of local ULIRGs.  Among the different
applications of these templates, we used them to  
 predict the intensities of the brightest
mid-IR emission lines for LIRGs as a function of redshift and
compare them with the expected sensitivities of future space IR
missions. The brightest mid-IR emission lines of LIRGs will be detected out
to $z\sim 1$, and may be used to constrain better the AGN and
star-formation properties of IR-bright galaxies.



$\,$

$\,$

The authors would  like to thank Takashi Hattori, Macarena
Garc\'{\i}a-Mar\'{\i}n, Bernhard Brandl,
Vassilis Charmandaris, Lee Armus, Guido Risaliti, 
Henrik Spoon, John Moustakas, and Duncan
Farrah  for their help and enlightening discussions, as well as 
Miwa Block for producing some of the data cubes. We are grateful
  to two anonymous referees for their constructive comments on the paper.

 This work was supported by NASA through contract 1255094 issued by
 JPL/California Institute of Technology.  AA-H, MP-S, LC, and TD-S 
acknowledge  support from the Spanish Plan Nacional del Espacio
through  grant
ESP2007-65475-C02-01. AA-H also acknowledges support from the
  Spanish Ministry of Science and Innovation through grant Proyecto Intramural
Especial 200850I003.

This research has made use of the NASA/IPAC Extragalactic Database (NED),
 which is operated by the Jet Propulsion Laboratory, California Institute of
 Technology, under contract with the National Aeronautics and Space
 Administration.

\end{document}